\documentclass[journal]{IEEEtran}
\usepackage{dsfont}
\usepackage{graphicx}
\usepackage{citesort}
\usepackage{bm}
\usepackage{amssymb}
\usepackage{amsfonts}
\usepackage{amsmath}
\usepackage{epsfig}
\usepackage{fancybox}
\usepackage{textcomp}
\usepackage{multirow}
\usepackage{booktabs}
\usepackage{mathrsfs}
\usepackage{pifont}

\usepackage[ruled,vlined]{algorithm2e}

\usepackage[dvipsnames,usenames]{color}
\usepackage[usenames,dvipsnames]{color}

    \newcommand{\qa}{{\bf a}}

    \newcommand{\qe}{{\bf e}}

    \newcommand{\qm}{{\bf m}}

    \newcommand{\qp}{{\bf p}}

    \newcommand{\qs}{{\bf s}}

    \newcommand{\qy}{{\bf y}}

    \newcommand{\qA}{{\bf A}}
    \newcommand{\qB}{{\bf B}}
    
    \newcommand{\qD}{{\bf D}}

    \newcommand{\qM}{{\bf M}}

    \newcommand{\qzero}{{\bf 0}}

    \newcommand{\qtheta}{{\boldsymbol \theta}}
    
    \newcommand{\qeta}{{\boldsymbol \eta}}
    
    \newcommand{\qmu}{{\boldsymbol \mu}}

    \newcommand{\qomega}{{\boldsymbol \omega}}
    \newcommand{\qsigma}{{\boldsymbol \sigma}}
    \newcommand{\qrho}{{\boldsymbol \rho}}

    \newcommand{\tqtheta}{\tilde{\boldsymbol \theta}}
    \newcommand{\hqtheta}{\hat{\boldsymbol \theta}}

    \newcommand{\bbR}{{\mathbb R}}

    \newcommand{\calE}{{\cal E}}
    \newcommand{\calF}{{\cal F}}

    \newcommand{\calL}{{\cal L}}
    
    \newcommand{\calN}{{\cal N}}

    \newcommand{\calU}{{\cal U}}

    \newcommand{\diag}{{\sf diag}}

    \newcommand{\Ex}{{\sf E}}

    \newcommand*{\argmin}{\operatornamewithlimits{argmin}\limits}
    \newcommand*{\argmax}{\operatornamewithlimits{argmax}\limits}

    \newcommand{\rme}{{\rm e}}
    \newcommand{\rms}{{\rm s}}
    \newcommand{\rmd}{{\rm d}}

    \newcommand{\sfb}{{\sf b}}
    
    \newcommand{\sfe}{{\sf e}}
    
    \newcommand{\sfo}{{\sf o}}
    \newcommand{\sfs}{{\sf s}}

    \newcommand{\sfP}{{\sf P}}
    \newcommand{\sfQ}{{\sf Q}}
    
    \newcommand{\sfZ}{{\sf Z}}

    \newcommand{\scrN}{\mathscr{N}}
    \newcommand{\scrL}{\mathscr{L}}

\begin{document}

\title{Location Identification of Power Line Outages \\ Using PMU
Measurements with Bad Data}

\author{
        Wen-Tai Li,
        Chao-Kai~Wen, 
        Jung-Chieh~Chen, 
        Kai-Kit~Wong, 
        Jen-Hao~Teng, %
        and~Chau~Yuen 

\thanks{W. T. Li, C. K. Wen, and J. H. Teng are with the Department of Electronic and Electrical Engineering, National Sun Yat-sen University, Kaohsiung 804, Taiwan. E-mail: {\rm chaokai.wen@mail.ee.nsysu.edu.tw}.}
\thanks{J. C. Chen is with the Department of Optoelectronics and Communication Engineering, National Kaohsiung Normal University, Kaohsiung 802, Taiwan. }
\thanks{K. Wong is with the Department of Electronic and Electrical Engineering, University College London, London, United Kingdom. }
\thanks{C. Yuen is with Engineering Product Development, Singapore University of Technology and Design, Singapore. }

}


\maketitle

\begin{abstract}
The use of phasor angle measurements provided by phasor measurement units (PMUs) in fault detection is regarded as a promising method in identifying locations of power line outages. However, communication errors or system malfunctions may introduce errors to the measurements and thus yield bad data. Most of the existing methods on line outage identification fail to consider such error. This paper develops a framework for identifying multiple power line outages based on the PMUs' measurements in the presence of bad data. In particular, we design an algorithm to identify locations of line outage and recover the faulty measurements simultaneously. The proposed algorithm does not require any prior information on the number of line outages and the noise variance. Case studies carried out on test systems of different sizes validate the effectiveness and efficiency of the proposed approach.
\end{abstract}

\section{Introduction}
\IEEEPARstart{P}{ower line} outage identification is of paramount importance for maintaining reliable and secure operation of electric power systems. When outages occur on power transmission lines, certain lines may become overloaded and consequently fail. Shortly thereafter, further cascading failures may result in system collapse. Therefore, a power system operator must accurately identify line outages promptly. Modern wide-area measurement system (WAMS), which builds upon phasor measurement units (PMUs) and fast communication links, is considered as a promising infrastructure for supporting fast line outage detection \cite{Aminifar-14Access}.

Several techniques for power line outage detection/identification based on PMUs' measurements have been investigated recently \cite{Tat-08,Tat-09,Sehwail-12,Emami-TPS13,He-TSG11,Abdelaziz-12,Zhu-12,Chen-TPS14,Zhao-S14,Wu-15TPS}. In particular, Tate and Overbye in \cite{Tat-08,Tat-09} proposed identification algorithms for single and double outage lines, respectively. The idea is to find a line combination so that the pre-computed phasor angle difference corresponding to that line outage event can match with the observed one. This methodology was further extended to accommodate islanding in \cite{Sehwail-12}. Zhu and Giannakis in \cite{Zhu-12} then used sparse configurations and proposed a compressed sensing based algorithm for identifying multiple line outages. Following \cite{Zhu-12}, Chen \emph{et al.}~\cite{Chen-TPS14} proposed an improvement method adopting a cross-entropy-based global optimization technique, and Zhao and Song \cite{Zhao-S14} presented a distributed framework to perform the identification locally at each phasor data concentrator. Most recently, Wu {\em et al.} \cite{Wu-15TPS} considered the same problem under scenarios with a limited number of PMUs.

These existing studies all rely on high accuracy of phasor angle measurements (or perfect PMUs). Although compared with traditional meters, PMUs are more robust against measurement errors, communication errors or system malfunctions may introduce errors to the measurements received by phasor data concentrators. In addition, it is very likely that certain physical impact on power system buses result in line outages and would subsequently lead to faulty PMUs. In these scenarios, a few of the phasor angle measurements may contain errors, which are referred to as {\em bad data}. Note that bad data are different from the common small additive noises resulting from certain uncertainties of PMUs (e.g., the A/D converters and instrument transformers). The uncertainties of PMUs are usually modeled as \emph{unstructured} noises, whose effects have been investigated in \cite{Chen-14JSTSP} for line outage identification, while bad data in the phasor angle vector lie in the range space of the susceptance matrix, which can arbitrarily perturb results of line outage identification.

Although the accuracy of line outage identification in the presence of bad data is expected to be degraded, a comprehensive study on this issue is missing. In this paper, we take the important first step to develop a framework for line outage identification based on phasor angle measurements with bad data. Our contribution is threefold:
\begin{itemize}
\item A line outage identification model is proposed with consideration of phasor angle measurements with bad data. Using this model, we not only can understand the influence of bad data on the identification problem but also can design a criterion to aid line outage detection. Particularly, location identification for line outage and bad data can be viewed as a sparse error detection problem, which permits us to identify them by leveraging on more recent techniques in compressive sensing\footnote{Compressive sensing is a signal processing technique for efficiently reconstructing a signal from an undersampled set of linear transformations.} \cite{Hayashi-13IEICE}.

\item We provide an effective algorithm for line outage identification. Unlike several of prior work (even without bad data), our scheme does not require prior information of the number of line outages and the noise variance. Specifically, all the required knowledge is learned as part of the identification procedure.

\item The developed algorithm operates in a message passing fashion, which greatly exploits the inherent \emph{sparsity} structure of power networks and thus leads to very low complexity. Comprehensive experimental studies show that the whole identification procedure can be completed in real-time even over a large number of bus systems (e.g., $\le 1$ second for a $2736$-bus system).
\end{itemize}


\section{System Model and Problem Formulation}\label{sec:02}
\subsection{DC Power Flow Model}
We consider a power transmission network with $N$ buses and $L$ transmission lines. Let $\scrN = \left\{1,\ldots,N\right\}$ be the set of buses and $\scrL = \left\{1,\ldots,L\right\}$ be the set of transmission lines. For the power transmission network, we adopt the most popular variant of the DC power flow model \cite{Wood-96BOOK}, in which the power flowing from buses $m$ to $n$ along line $l \in \scrL$ can be presented as
\begin{equation} \label{eq:powerFlow_nm}
    p_{nm} = \frac{1}{x_{nm}} (\theta_n-\theta_m),
\end{equation}
where ${x_{nm} = x_{mn}}$ represents the reactance between buses $n$ and $m$, and $\theta_n$ and $\theta_m$ are their respective voltage phasor angles.

Let $p_n$ be the nodal injection for bus $n$. The nodal flow conservation constraint state that the amount of power injected into bus $n$ must be equal to the amount that flows out of it, which can be algebraically expressed as
\begin{equation}\label{eq:1}
    p_n=\sum_{m\in \scrN(n)}p_{nm},
\end{equation}
where $\scrN(n)$ denotes the set of neighboring buses connected to bus $n$. Then (\ref{eq:1}) together with (\ref{eq:powerFlow_nm}) yields the following linear DC power flow model in matrix form
\begin{equation}\label{eq:DC-power-flow-model}
    \qp = \qB \qtheta,
\end{equation}
where ${\qp = [p_1 \cdots p_N ]^T} \in \mathbb{R}^{N}$, ${\qtheta= [\theta_1 \cdots \theta_N ]^T} \in \mathbb{R}^{N}$, and $\qB = [B_{nm}] \in \mathbb{R}^{N \times N}$ with its $(n,m)$th entry given by $B_{nm}=-\frac{1}{x_{nm}}$ if $m\in \scrN(n)$ and $m \neq n$, $B_{nm}=\sum_{m\in \scrN(n)} \frac{1}{x_{nm}}$ if $n=m$, and $B_{nm}=0$ otherwise.


Recall that line $l$ connects buses $n$ and $m$. If we define the $i$-th element of line $l$ \emph{incidence} vector $\qm_l$ as
\begin{equation}\label{eq:incidence_vector}
    \qm_l\left(i\right)=\left\{
    \begin{array}{rl}
                  1,  & \mbox{if $i=n$},\\
                  -1, & \mbox{if $i=m$}, \\
                  0,  & \mbox{otherwise},
    \end{array}\right.
\end{equation}
then $\qB$ in (\ref{eq:DC-power-flow-model}) can be expressed as \cite{Zhu-12}
\begin{equation}\label{eq:B_def}
    \qB = \qM\qD_{x}\qM^T = \sum_{l=1}^{L}\frac{1}{x_{l}} \qm_l \qm_l^{T},
\end{equation}
where $\qD_x$ is a diagonal matrix with $x_{l}^{-1}$ as its $l$-th diagonal entry, and $\qM = [\qm_1 \cdots \qm_L ] $ is the $N \times L$ bus-line incidence matrix.

\subsection{Power Line Outages}
From (\ref{eq:DC-power-flow-model}), we see that the relationship between the injected power vector $\qp$ and the pre-event phasor angle vector $\qtheta$ is dictated by the susceptance matrix $\qB$ which is topology-dependent. Following \cite{Tat-08,Tat-09}, we assume that the post-outage grid remains connected when outages occur on the transmission lines. As the interconnected grid have reached a stable post-event state, the post-event power flow can be expressed by
\begin{equation}\label{eq:post-event-DC-power-flow-model}
    \qp'= \qB'\qtheta' = \qp + \qeta,
\end{equation}
where $\qB'$ and $\qtheta'$ are the post-event susceptance matrix and the post-event phasor angle vector, respectively, and $\qeta$ denotes the small perturbations between $\qp'$ and $\qp$, usually modeled as a Gaussian noise vector with zero mean and covariance matrix $\sigma^{2}_{\eta}\,\bf{I}$ \cite{Sch-05}.

To reflect variations in the post-event, we write
\begin{equation}
    \qB' = \qB - \Delta\qB ~~~~\mbox{and}~~~~
    \qtheta' = \qtheta + \Delta\qtheta, \label{eq:deltaBATheta_def}
\end{equation}
where $\Delta\qB$ and $\Delta\qtheta$ represent variations of the susceptance matrix and the phasor angle vector, respectively, between the pre- and post-event power systems. Using the notations of (\ref{eq:B_def}), $\Delta\qB$ can be expressed as
\begin{align}
    \Delta\qB & = \sum_{l \in \calL_{\sfo}}\frac{1}{x_l}\qm_l \qm_l^T
    = \qM \qD_x \diag\left(\qs_{\sfo}\right) \qM^T \label{eq:tilde_B},
\end{align}
where $\calL_{\sfo} \subset \scrL$ denotes the set of the lines in outage, and $\qs_{\sfo} = [s_{\sfo,1} \dots s_{\sfo,L} ]^T $ is an $L$-dimensional \emph{binary} vector whose element $s_{\sfo,l} = 1$ if the $l$-th line belongs to $\calL_{\sfo}$ and $s_{\sfo,l} = 0$ otherwise.

Substituting (\ref{eq:deltaBATheta_def}) into (\ref{eq:post-event-DC-power-flow-model}) yields
\begin{align}
    \qy \triangleq \qB \Delta\qtheta
    &= \Delta\qB\qtheta' + \qeta \label{eq:def_y} \\
    &=  \qM \qD_x {\diag\left( \qM^T\qtheta'\right)}\qs_{\sfo}+\qeta, \label{eq:def_y2}
\end{align}
where the last equality follows from the fact that ${\diag\left(\qs_{\sfo}\right)} \qM^T \qtheta' = {\diag\left(\qM^T \qtheta'\right)} \qs_{\sfo}$. By introducing the notation
\begin{equation} \label{eq:defA}
    \qA_{\qtheta'} = \qM \qD_x {\diag\left( \qM^T\qtheta'\right)},
\end{equation}
we then arrive at
\begin{equation}\label{eg:P1-y}
 \qy  = \qA_{\qtheta'}\qs_{\sfo}+\qeta.
\end{equation}
Here, the notation $\qA_{\qtheta'} \in \bbR^{N \times L}$ indicates that matrix $\qA_{\qtheta'}$ depends on $\qtheta'$. Note that since $\qB$ and $\Delta\qtheta$ are available, $\qy$ can be obtained by its definition in (\ref{eq:def_y}). In addition, since the pre-event network topology (i.e., $\qM$ and $\qD_x$) as well as the the post-event phasor angle vector $\qtheta'$ are known, $\qA_{\qtheta'}$ is also available by (\ref{eq:defA}). Therefore, with (\ref{eg:P1-y}), the power line outages can be identified by solving
\begin{equation}\label{problem:1}
    \textsf{P1}:~~ \widehat{\qs}_{\sfo} =\argmin_{\qs_{\sfo} \in \{0,1\}^L} {\calF_1\left(\qs_{\sfo}; \qy, \qA_{\qtheta'} \right)},
\end{equation}
where $\calF_1$ is a cost function of $\qs_{\sfo}$ associated with the model in (\ref{eg:P1-y}). For example, $\calF_1\left(\qs_{\sfo}; \qy, \qA_{\qtheta'} \right) = \left\|{\bf y}-\qA_{\qtheta'}\qs_{\sfo}\right\|_2^2$ is adopted in \cite{Tat-08,Tat-09,Zhu-12,Chen-TPS14} for line outage identification applications.

\subsection{Power Line Outages with Bad Data}
When the phasor angle measurements $(\qtheta,\qtheta')$ are accurate, recent reports \cite{Tat-08,Tat-09,Zhu-12,Chen-TPS14} have verified the efficacy of Problem \textsf{P1} for line outage identification. However, if some measurements of $(\qtheta,\qtheta')$ are erroneous or bad, Problem \textsf{P1} will result in incorrect line outage identification. To better understand the line outage identification problem with bad data, we aim to develop a corresponding line outage identification model of (\ref{eg:P1-y}) while some bad datums are present in $\qtheta'$.\footnote{Bad data could also be present in $\qtheta$. Since we only utilize the differences between the pre- and post-event measurements, the bad data in $\qtheta$ can be included in $\qtheta'$.}  We denote the corrupted measurement of $\qtheta'$ by
\begin{equation} \label{eq:thetaWb_def}
    \tqtheta' = \qtheta' + \qtheta_{\sfb}',
\end{equation}
where $\qtheta_{\sfb}'$ is an unknown vector with its $n$-th entry being nonzero only if the entry is a bad datum. Similar to (\ref{eq:deltaBATheta_def}), we let
\begin{equation} \label{eq:deltaThetaWb_def}
    \tqtheta'= \qtheta + \Delta\qtheta_{\sfb}
\end{equation}
be the post-event phasor angle vector. Note that $\Delta\qtheta_{\sfb}$ contains not only variations of the phasor angle vector between the pre- and post-event power systems but also the bad data.

In this case, $\qy$ in (\ref{eq:def_y}) becomes
\begin{equation}
    \qy = \qB \Delta\qtheta_{\sfb}
    = \qB (\Delta\qtheta + \qtheta_{\sfb}')
    = \Delta\qB \qtheta' + \qB\qtheta_{\sfb}' + \qeta, \label{eq:y_badData1}
\end{equation}
where the second equality follows by simply substituting the definitions in (\ref{eq:deltaBATheta_def}), (\ref{eq:thetaWb_def}), and (\ref{eq:deltaThetaWb_def}), and the last equality follows by the equality in (\ref{eq:def_y}). Comparing (\ref{eq:y_badData1}) with (\ref{eq:def_y2}), we see that $\qB\qtheta_{\sfb}'$ in (\ref{eq:y_badData1}) is the effect due to the bad data. Recalling from (\ref{eq:thetaWb_def}), $\qtheta_{\sfb}'$ is a sparse vector. To clarify the effect due to the bad data, we let $\calL_{\sfb} \subset \scrL$ be the set of lines, whose element are those lines connected to the faulty buses. Also, let $\qs_{\sfb}$ be an $L$-dimensional \emph{binary} vector whose element $s_{\sfb,l} =1$ if the $l$-th line belongs to $\calL_{\sfb}$ and $s_{\sfb,l} = 0$ otherwise. Thus, we use (\ref{eq:deltaBATheta_def}) and these notations to write
\begin{equation} \label{eg:Btheta_badData}
  \qB\qtheta_{\sfb}' = (\Delta\qB+\qB') \qtheta_{\sfb}' = (\Delta\qB+\Delta\qB_{\sfb}) \qtheta_{\sfb}',
\end{equation}
where $\Delta\qB_{\sfb} \triangleq \qM \qD_{x}\diag(\qs_{\sfb})\qM^T$.\footnote{We notice that $\Delta\qB \qtheta_{\sfb}' = \qzero$ if $\calL_{\sfo} \cap\calL_{\sfb} = \emptyset$ and $\Delta\qB \qtheta_{\sfb}' \neq \qzero$ otherwise.}

\begin{figure}
\begin{center}
\resizebox{2.75in}{!}{%
\includegraphics*{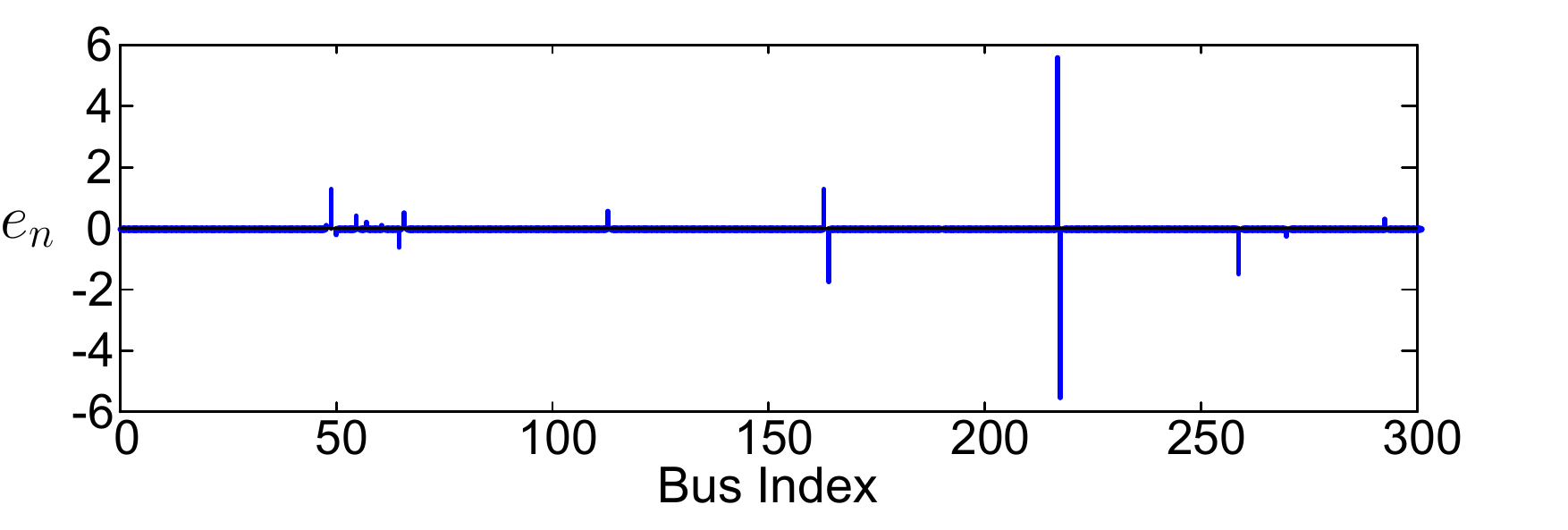} }%
\caption{A realization of $\qe$ for a 300-bus system with $5$ faulty buses.}\label{fig:Example_e}
\end{center}
\end{figure}

Recalling the definition of $\qA$ from (\ref{eq:defA}) and using (\ref{eq:thetaWb_def}), we write
\begin{equation} \label{eq:defAtqtheta}
    \qA_{\tqtheta'} = \qM \qD_x {\diag( \qM^T\tqtheta')}
    = \qA_{\qtheta'} + \qA_{\qtheta_{\sfb}'}.
\end{equation}
Then substituting (\ref{eg:Btheta_badData}) into (\ref{eq:y_badData1}) shows that
\begin{align}
    \qy 
    &= \Delta\qB(\qtheta' + \qtheta_{\sfb}') + \Delta\qB_{\sfb}\qtheta_{\sfb}'  + \qeta \nonumber \\
    &= \qA_{\tqtheta'} \qs_{\sfo} + \qA_{\qtheta_{\sfb}'} \qs_{\sfb} + \qeta \nonumber \\
    &= \qA_{\tqtheta'} (\qs_{\sfo} + \qs_{\sfb}) - \qA_{\qtheta'} \qs_{\sfb} + \qeta, \label{eq:y_badData}
\end{align}
where the second equality follows the similar algebraic step in (\ref{eq:def_y2}), and the last equality follows by (\ref{eq:defAtqtheta}). By introducing
\begin{equation} \label{eq:def_sAe}
    \qs \triangleq \qs_{\sfo} + \qs_{\sfb} ~~~\mbox{and}~~~
    \qe \triangleq -\qA_{\qtheta'} \qs_{\sfb},
\end{equation}
we thus arrive at
\begin{equation} \label{eg:y_badData3}
 \qy  = \qA_{\tqtheta'}\qs + \qe + \qeta.
\end{equation}
Note that $\qs$ is still a \emph{binary} vector. In addition, since matrix $\qA_{\qtheta'}$ and vector $\qs_{\sfb}$ are sparse, $\qe \in \bbR^{N} $ is also a sparse vector, which results in \emph{sparse} contamination on $\qy$. See an example of $\qe$ in Figure \ref{fig:Example_e}.

\begin{figure}
\begin{center}
\resizebox{3.25in}{!}{%
\includegraphics*{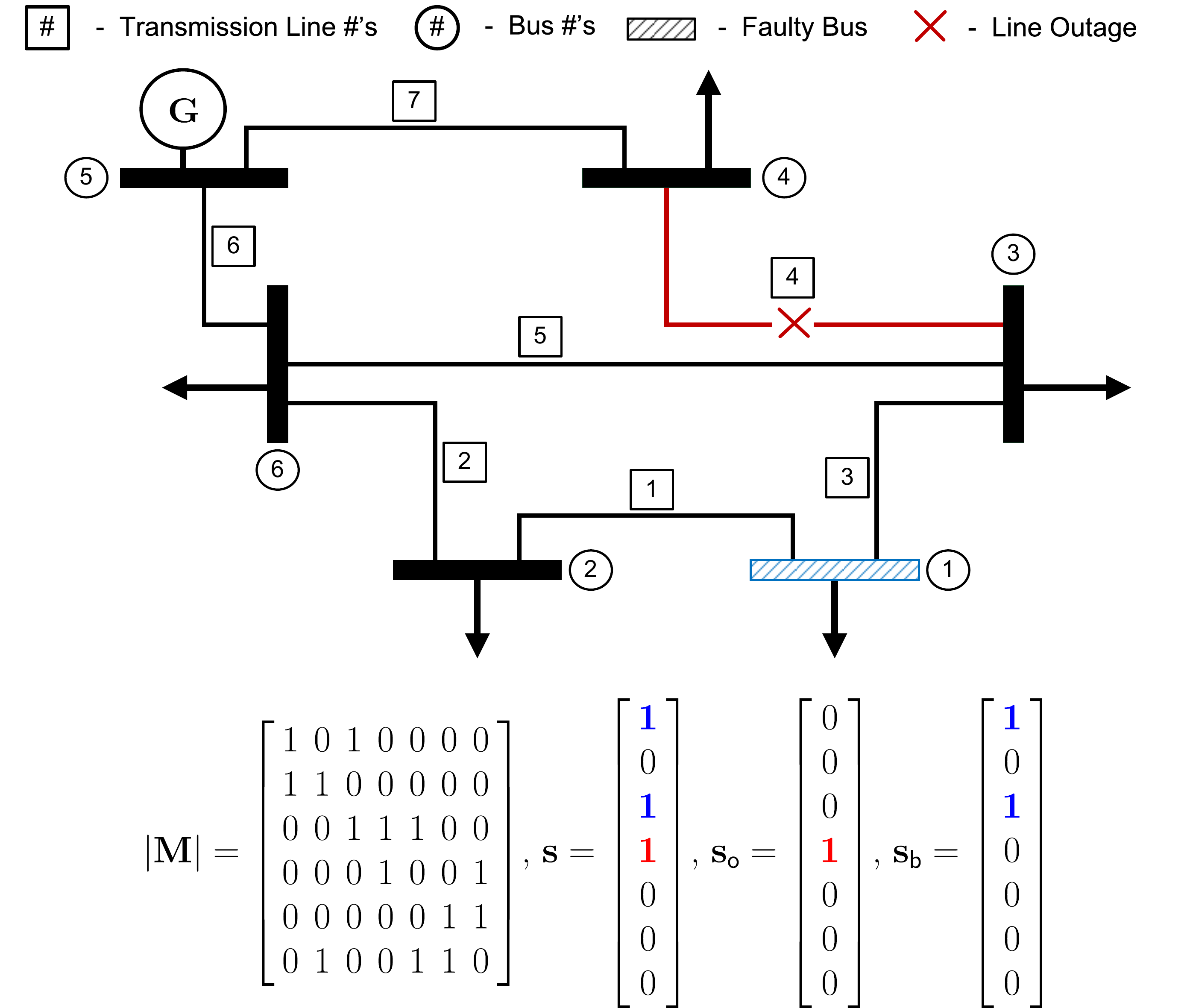} }%
\caption{A six-bus system.}\label{fig:Baddata_example}
\end{center}
\end{figure}

According to the discussion in Section II-B, matrix $\qA_{\tqtheta'}$ is available while $\qs$, $\qe$, and $\qeta$ are unknown. Clearly from (\ref{eg:y_badData3}), as the bad data are present, the line outages cannot be effectively identified using Problem \textsf{P1} in (\ref{problem:1}). This is not only because the additional $\qe \in \bbR^{N}$ contaminates $\qy$ but also because both $\qs_{\sfo}$ and $\qs_{\sfb}$ are involved in $\qs$. To address this problem, we propose to estimate both $\qs$ and $\qe$ from $\qy$ by solving the following optimization problem:
\begin{equation}\label{problem:2}
    \textsf{P2}:~~ (\widehat{\qs},\widehat{\qe})
    =\argmin_{\qs \in \{0,1\}^L, \qe \in \bbR^{N}}
    {\calF_2\left(\qs, \qe; \qy, \qA_{\tqtheta'} \right)},
\end{equation}
where $\calF_2$ is a cost function of $(\qs, \qe)$ associated with the model in (\ref{eg:y_badData3}). However, note that even though $\qs$ can be successfully estimated via Problem \textsf{P2}, the locations of the line outages are still unknown. As mentioned previously, $\qs$ defined in (\ref{eq:def_sAe}) contains the location information of both the line outages and the bad data. In subsequent sections, we first provide a way to separate $\qs_{\sfo}$ from $\qs$ in Section \ref{sec:03}, and postpone solving Problem \textsf{P2} to Section \ref{sec:04}.

\section{Line Outage Identification with Bad Data}\label{sec:03}
To start with, we first assume that $\qs$ has been obtained successfully from Problem \textsf{P2}, i.e., $\widehat{\qs} = \qs_{\sfo} + \qs_{\sfb}$. Before proceeding, we make the following definitions for ease of exposition. Recall the sets of \emph{lines} $\calL_{\sfo}$ and $\calL_{\sfb}$, whose elements are the lines in outage and the lines connected to the faulty buses, respectively. Let $\calN_{\sfo}$ and $\calN_{\sfb}$ be the sets of \emph{buses} associated with $\calL_{\sfo}$ and $\calL_{\sfb}$, respectively. Specifically, set $\calN_{\sfo}$ contains all the buses involving the line outages $\calL_{\sfo}$, and set $\calN_{\sfb}$ contains all the buses connected by the lines of $\calL_{\sfb}$. Let $\calE_{\sfb}$ be the set of the faulty buses. Since each line connects to two buses, $\calE_{\sfb}$ is only a subset of, but \emph{not} equal to, $\calN_{\sfb}$. For a better understanding on these definitions, we provide an example as shown in Figure \ref{fig:Baddata_example}, where line~$4$ is in outage and bus~$1$ is faulty. Therefore, $\calL_{\sfo} = \{ 4 \}$, $\calN_{\sfo} = \{ 3,4 \}$, $\calE_{\sfb} =\{ 1\}$, $\calL_{\sfb} = \{ 1,3 \}$, and $\calN_{\sfb} = \{ 1,2,3 \}$.

Next, we provide the ways to 1) separate $\widehat{\qs}_{\sfo}$ from $\widehat{\qs}$ and 2) recover $\qtheta'$ from its corrupted measurement $\tqtheta'$, which are referred to as the separation phase (or {\sf S-phase}) and the recovering phase (or {\sf R-phase}), respectively.

{\bf S-Phase}---Recall the bus-line incidence matrix $\qM$ from (\ref{eq:incidence_vector}) and (\ref{eq:B_def}); see also Figure \ref{fig:Baddata_example} for an example. If we assume that there are \emph{at most} one line in outage on each bus, then set $\calN_{\sfo}$ can be included in $\{ {n \in \scrN} : \sum_{l=1}^{L}|M_{nl}| \widehat{s}_{l} = 1\}$. Therefore, we can separate the faulty buses from $\widehat{\qs}$ by
\begin{equation} \label{eq:defEbset}
 \widehat{\calE}_{\sfb} \triangleq \Bigg\{ {n \in \scrN} : \sum_{l=1}^{L} |M_{nl}| \widehat{s}_{l} > 1 \Bigg\}.
\end{equation}
Clearly, if $\widehat{\calE}_{\sfb} = \emptyset$, it means that no bad data is present. With $\widehat{\calE}_{\sfb}$, we further define
\begin{equation} \label{eq:defLbset}
 \widehat{\calL}_{\sfb} \triangleq \Bigg\{ {l \in \scrL} : \sum_{n \in \widehat{\calE}_{\sfb}} |M_{nl}| > 0 \Bigg\},
\end{equation}
which thus induces the following set
\begin{equation} \label{eq:defNbset}
 \widehat{\calN}_{\sfb} \triangleq \Bigg\{ {n \in \scrN} : \sum_{l \in \widehat{\calL}_{\sfb}} |M_{nl}| > 0 \Bigg\}.
\end{equation}
We can see that $\widehat{\calE}_{\sfb} = \calE_{\sfb}$, $\widehat{\calL}_{\sfb} = \calL_{\sfb}$ and $\widehat{\calN}_{\sfb} = \calN_{\sfb}$. These relations can be easily understood through the example in Figure \ref{fig:Baddata_example}. With $\widehat{\calL}_{\sfb}$, we can determine $\widehat{\qs}_{\sfb}$ (the estimate of $\qs_{\sfb}$) by setting $\widehat{s}_{\sfb,l} = 1$ if the $l$-th line belongs to $\widehat{\calL}_{\sfb}$ and $\widehat{s}_{\sfb,l} = 0$ otherwise. Eventually, we obtain ${\widehat{\qs}_{\sfo} = \widehat{\qs} - \widehat{\qs}_{\sfb}}$ (the estimate of $\qs_{\sfo}$), and complete the {\sf S-phase}.

However, notice that the above argument is based on the assumption of at most one line outage on each bus. If this constraint is relaxed, $\widehat{\calE}_{\sfb}$ shall contain some of buses connected by such outage lines. It turns out that $\widehat{\calN}_{\sfb}$ shall contain some buses in $\calN_{\sfo}$. Fortunately, this confusion can be eliminated in the subsequent {\sf R-phase}.

{\bf R-Phase}---The aim of this phase is to recover $\qtheta'$ from its corrupted measurement $\tqtheta'$. Letting $\qy_{\sfb} = \qA_{\tqtheta'}\widehat{\qs} -\qy$, we then use (\ref{eq:y_badData}) to write
\begin{equation}
 \qy_{\sfb} = \qA_{\qtheta'} \widehat{\qs}_{\sfb} - \qeta
  = \qM \qD_x {\diag(\widehat{\qs}_{\sfb})} \qM^T \qtheta' - \qeta, \label{eq:defyb2}
\end{equation}
where the second equality follows by the similar algebraic step in (\ref{eq:def_y2}). Notice that not all the entries of $\qtheta'$ should be estimated. Only a few \emph{corrupted} measurements of $\qtheta'$, whose locations have been identified by $\widehat{\calE}_{\sfb}$, should be recovered.

Toward this end, we first make the following definitions. For any vector $\qa \in \bbR^{N}$ and index set $\alpha \subseteq \{ 1, \ldots, N \}$, we denote the (sub)vector that lies in the entries of $\qa$ indexed by $\alpha$ as $[\qa]_{\alpha}$. Similarly, for any matrix $\qA \in \bbR^{N \times N}$, we denote the (sub)matrix that lies in the rows and columns of $\qA$ indexed by $\alpha$ as $[\qA]_{\alpha}$. The cardinality $|\alpha|$ denotes the number of members of $\alpha$. Then from (\ref{eq:defyb2}), we find that recovering $[\hqtheta']_{\widehat{\calE}_{\sfb}}$ (the corrupted measurements of $\qtheta'$) is possible through solving the following optimization problem:
\begin{align} \label{eq:Problem3}
 \hspace{-0.25cm} & \textsf{P3}:~ \big[\hqtheta'\big]_{\widehat{\calN}_{\sfb}}
 \nonumber \\
 \hspace{-0.25cm} &~~~~~~~~ = \argmin_{\left[\qtheta'\right]_{\widehat{\calN}_{\sfb}}} ~ \left\| \big[ \qy_{\sfb} \big]_{\widehat{\calN}_{\sfb}} - \big[\qM \qD_x \diag(\widehat{\qs}_{\sfb}) \qM^T \big]_{\widehat{\calN}_{\sfb}} \big[\qtheta'\big]_{\widehat{\calN}_{\sfb}} \right\|_2^2  \nonumber \\
 \hspace{-0.25cm} &~~~~~~~~ \quad\quad \mbox{s.t.} \qquad \theta'_{n} = \tilde{\theta}'_{n}, ~\forall\,n \neq \widehat{\calE}_{\sfb}.
\end{align}
Here, the estimate of $\qtheta'$ is denoted by $\hqtheta'$.
Since $\widehat{\calE}_{\sfb} \subset \widehat{\calN}_{\sfb}$, the estimate $[\hqtheta']_{\widehat{\calN}_{\sfb}}$ has involved the estimate $[\hqtheta']_{\widehat{\calE}_{\sfb}}$. Problem \textsf{P3} can be easily solved by eliminating the known variables $\{ \theta'_{n} = \tilde{\theta}'_{n}, \,\forall n \neq \widehat{\calE}_{\sfb} \}$ from its objective function, and then applying the linear least square method to solve the unknown variables.

If $\widehat{\calE}_{\sfb} = \calE_{\sfb}$, the above procedure has recovered $[\hqtheta']_{\widehat{\calE}_{\sfb}}$ from the corrupted measurements. However, as mentioned in the {\sf S-phase}, if there are more than one line outages on a bus, such a bus cannot be identified as $\calN_{\sfo}$, but is included in $\widehat{\calN}_{\sfb}$. In this case, some of lines in $\widehat{\calL}_{\sfb}$ should be in the outage state. That is, the zero-one state of $[\widehat{\qs}_{\sfb}]_{\widehat{\calL}_{\sfb}}$ is uncertain rather than $[\widehat{\qs}_{\sfb}]_{\widehat{\calL}_{\sfb}} = {\bf 1}$ as given in the {\sf S-phase}. To determine its state, an exhaustive search (ES) algorithm is employed to evaluate all possible combinations $\big[ \widehat{\qs}_{\sfb} \big]_{\widehat{\calL}_{\sfb}} \in \{ 0, 1\}^{|\widehat{\calL}_{\sfb}|}$, and then find the combination that yields the minimum error of Problem \textsf{P3}. As $|\widehat{\calL}_{\sfb}|$ is very small (e.g., $|\widehat{\calL}_{\sfb}| = 2$ in Figure \ref{fig:Baddata_example}) and does not expand with the number of buses, ES for phase recovering can be realized in real time. Consequently, we have simultaneously identified $\qs_{\sfb}$ and recovered $\qtheta'$ and therefore completed the {\sf R-phase}.

The step-wise implementation procedure of the proposed line outage identification algorithm is summarized as Algorithm \ref{ago:agoIdAndRePhase}. In short, we first obtain $(\widehat{\qs},\widehat{\qe})$ by solving Problem \textsf{P2} (lines 1--2 of Algorithm \ref{ago:agoIdAndRePhase}). Next, $\widehat{\qs}_{\sfb}$ is separated from $\widehat{\qs}$ via the {\sf S-phase}  (lines 3--4) and then refined by the {\sf R-phase} (lines 5--6). The locations of line outages are finally indicated by $\widehat{\qs}_{\sfo}$ at line 7.

\begin{algorithm}[!h]\label{ago:agoIdAndRePhase}  \footnotesize
  \caption{Line Outage Identification with Bad Data}
  \SetKwInOut{Input}{input}
  \SetKwInOut{Output}{output}
  \SetKwProg{Fn}{}{\string:}{}

  \Input{The pre-event phasor angle vector $\qtheta$, the post-event phasor angle vector $\tqtheta'$, and the the pre-event susceptance matrix $\qB$ }
  \Output{The indicator vector for line outages $\widehat{\qs}_{\sfo}$ }
  \BlankLine
  \Begin{
  \nl Generate $\qy = \qB (\tqtheta'-\qtheta)$\;
  \nl Estimate $(\widehat{\qs},\widehat{\qe})$ by using Problem \textsf{P2} in (\ref{problem:2})\;
  \Fn{{\rm \bf Separation Phase}}{
  \nl Find sets $\widehat{\calE}_{\sfb}$, $\widehat{\calL}_{\sfb}$, $\widehat{\calN}_{\sfb}$ by using (\ref{eq:defEbset}), (\ref{eq:defLbset}), and (\ref{eq:defNbset}), respectively\;
  \nl Get $\widehat{\qs}_{\sfb}$ from $\widehat{\calL}_{\sfb}$\;
  }
  \Fn{{\rm \bf Recovering Phase}}{
  \nl Generate $\qy_{\sfb} = \qA_{\tqtheta'}\widehat{\qs} - \qy$\;
  \nl Estimate $[\hqtheta']_{\widehat{\calN}_{\sfb}}$ and refine $[ \widehat{\qs}_{\sfb} ]_{\widehat{\calL}_{\sfb}}$ simultaneously by using Problem \textsf{P3} in (\ref{eq:Problem3})\;
  }
  \nl Return $\widehat{\qs}_{\sfo} = \widehat{\qs}-\widehat{\qs}_{\sfb}$.
  }
\end{algorithm}

\section{Estimation Algorithm}\label{sec:04}
Now, we consider the estimation of $(\qs,\qe)$ based on Problem \textsf{P2} in (\ref{problem:2}). This task seems rather impossible because the total number of unknown variables ${L + N}$ is much larger than the number of observations $N$. Nevertheless, it is noted that $\qs$ and $\qe$ are sparse vectors (see the discussion in Section \ref{sec:02}-C). By exploiting the sparsity property of $(\qs,\qe)$, we can estimate them accurately by leveraging on recent techniques in compressive sensing (CS in brief in the sequel), see \cite{Hayashi-13IEICE} for a recent exhaustive list of the algorithms.

In the CS literature, one popular suboptimal and low-complexity estimator is $\ell_1$-regularized least-squares (LS), a.k.a.~least absolute shrinkage and selection operator (LASSO) \cite{Tibshirani-96JRSS}. In this context, the cost function of Problem \textsf{P2} is given by
\begin{equation} \label{eq:costFun_lasso}
    {\calF_2\left(\qs, \qe; \qy, \qA \right)}
    \\=  \left\|{\bf y}-\qA\qs - \qe \right\|_2^2 + \lambda_s \| \qs \|_1 + \lambda_e \| \qe \|_1 ,
\end{equation}
where $\lambda_s,\lambda_e > 0$ are the regularization parameters. We here and hereafter denote ${\qA := \qA_{\tqtheta'}}$ when it is not useful to specify the dependence on $\tqtheta'$ for matrix $\qA$. If the phasor angle measurements are accurate (i.e., $\qe = \qzero$), (\ref{eq:costFun_lasso}) reduce to ${\calF_2\left(\qs; \qy, \qA \right)} =  \left\|{\bf y}-\qA\qs \right\|_2^2 + \lambda_s \| \qs \|_1$. This cost function is adopted by \cite{Zhu-12} for the line outage identification problem \emph{without} bad data. It is known that large values of the regularization parameters result in more sparsity in $(\widehat{\qs}, \widehat{\qe})$. However, the best choice of $(\lambda_s,\lambda_e)$ highly depends on the statistical properties of $(\qs, \qe)$ (e.g., the sparsity of $(\qs, \qe)$) and the noise variance $\sigma_{\eta}^2$ \cite{Niazadeh-10}, which could be difficult to determine in practice. In addition, LASSO is highly suboptimal and thus would  not be quite suitable for the power line outage identification problem which requires very high reliability. The remainder of this section is devoted to devising a fast near-optimal algorithm for estimating $(\qs,\qe)$ from $\qy$.

\subsection{Theoretical Foundation}\label{sec:04-1}
To develop our algorithm, we adopt the probabilistic Bayesian inference because this approach provides a foundation for achieving the best estimates in terms of mean-squared error \cite{Poor-94BOOK}. Most importantly, the Bayesian inference can be implemented by a factor-graph framework which leads to low-complexity message-passing solutions.

Bayesian inference begins with deriving the posterior probability according to Bayes' rule:
\begin{equation} \label{eq:postProbability}
    \sfP(\qs,\qe|\qy) = \frac{\sfP(\qy|\qs,\qe)\sfP_{\sfs}(\qs)\sfP_{\sfe}(\qe)}{\sfP(\qy)},
\end{equation}
where $\sfP(\qs)$ and $\sfP(\qe)$ are the prior distributions of $\qs$ and $\qe$, respectively,  $\sfP(\qy|\qs,\qe)$ is the likelihood, and $\sfP(\qy)$ is the marginal likelihood. Specifically, the likelihood derived from the conditional distribution of $\qy$ based on (\ref{eg:y_badData3}) is given by
\begin{equation} \label{eq:condProbability}
    \sfP(\qy|\qs,\qe) =  \frac{1}{(2 \pi \sigma_{\eta}^2)^{\frac{N}{2}}}
    e^{-\frac{1}{2 \sigma_{\eta}^2 } \|\qy - \qA \qs - \qe \|_2^2}.
\end{equation}

With $\sfP(\qs,\qe|\qy)$, the marginal posterior probabilities of $\qs$ and $\qe$ can be obtained by ${\sfP(\qs|\qy) = \int \sfP(\qs,\qe|\qy) d\qe}$ and $\sfP(\qe|\qy) = \sum_{\qs \in \{ 0,1\}^N} \sfP(\qs,\qe|\qy) $, respectively. Then the Bayes-optimal way to estimate $\qs$ and $\qe$ is given by \cite{Poor-94BOOK}
\begin{equation} \label{eq:est_sAe}
     \widehat{s}_{l} = {\sum_{s_l \in \{ 0,1\} } s_{l} \, \sfQ(s_{l})} \mbox{~~and~~}
     \widehat{e}_{n} = {\int e_{n} \sfQ(e_{n}) \rmd e_{n}},
\end{equation}
where
\begin{equation} \label{eq:Qmarginal}
    \sfQ(s_{l}) = \sum_{\qs_{\backslash l} \in \{ 0,1\}^{L-1}}  \sfP(\qs|\qy) \mbox{~~and~~}
    \sfQ(e_{n}) = \int \sfP(\qe|\qy) \rmd \qe_{\backslash n}
\end{equation}
denote the marginal posterior probabilities of $s_{l}$ and $e_{n}$, respectively. Here, notation $\alpha_{\backslash i}$ stands for the set of all entries in $\alpha$ except for the entry indexed by $i$; for example, $\qs_{\backslash l} = [s_{1} \cdots s_{l-1} \, s_{l+1} \cdots s_{L} ]^T$ and $\scrN_{\setminus n} = \{1, \dots, {n-1}, {n+1}, \dots, N \}$.

From (\ref{eq:postProbability}), to obtain the posterior probability, the prior distributions of $\qs$ and $\qe$ are required. For line outages, it is reasonable to assume them to be independent and identically distributed (i.i.d.) random variables with Bernoulli distribution
\begin{equation} \label{eq:P_s1}
    {\sfP_{\sfs}({s_{l}=1};p_{\sfo})} = p_{\sfo} = 1-{\sfP_{\sfs}({s_{l}=0};p_{\sfo})}.
\end{equation}
Then the prior probability of $\qs$ can be expressed as
\begin{equation} \label{eq:P_s}
    \sfP_{\sfs}(\qs;p_{\sfo}) = \prod_{l=1}^{L} \sfP_{\sfs}(s_l;p_{\sfo}).
\end{equation}
Also, from Figure \ref{fig:Example_e}, we see that $\qe$ consists of sparse impulsive components, and the impulsive components have significantly different variances. These observations motivate us to model the elements of $\qe=[e_{n}] $ by a Bernoulli-Gaussian-mixture (B-GM) distribution:
\begin{equation} \label{eq:P_en}
    \sfP_{\sfe}(e_{n}; \qrho,\qmu,\qsigma^2) = \rho_{0}\delta(e_{n}) + \sum_{k=1}^{K} \rho_{k} \calN(e_{n}; \mu_k, \sigma_{k}^2),
\end{equation}
where $\delta(\cdot)$ denotes the Dirac delta, $\calN(e_{n}; \mu_k, \sigma_k^2)$ denotes a Gaussian probability density function (pdf) with mean $\mu_k$ and variance $\sigma_{k}^2$, $ \rho_{k}$ is the mixing probability of the $k$th GM component, and $\sum_{k=0}^{K} \rho_{k}  = 1$. The value of $K$ indicates the number of different variances in $\qe$. In the simulations of Section V, we use $K = 3$. Letting ${\qomega \triangleq (\qrho,\qmu,\qsigma^2)}$, the prior probability of $\qe$ is written as
\begin{equation} \label{eq:P_e}
    \sfP_{\sfe}(\qe; \qomega) = \prod_{n=1}^{N} \sfP_{\sfe}(e_{n}; \qomega).
\end{equation}
Note that the true distributions of $\qe$ could not be the B-GM distribution. However, our numerical results will demonstrate that the choice of the B-GM distribution is perfectly fine.

Even with the probability models of $\qs$ and $\qe$, there are two critical issues when implementing the optimal Bayes estimation (\ref{eq:est_sAe}). First,  the marginal posterior probabilities $\sfQ(s_{l})$ and $\sfQ(e_{n})$ in (\ref{eq:Qmarginal}) are not computationally tractable. Second, the prior parameters $(p_{\sfo},\qomega)$ for $(\sfP_{\sfs},\sfP_{\sfe})$ are unknown. To obtain an estimate of $\{\sfQ(s_{l}), \sfQ(e_{n})\}$, we use belief-propagation (BP) which is an iterative message passing algorithm in \cite{Donoho-09PNAS,Krzakala-12JSM}. Meanwhile, we use the expectation-maximization (EM) algorithm in \cite{Vila-13SP} to learn the prior parameters $(p_{\sfo},\qomega)$. We describe the two algorithms and their connections next.

\subsection{Message Passing Algorithm}\label{sec:04-2}

\begin{figure}
\begin{center}
\resizebox{!}{2.75in}{%
\includegraphics*{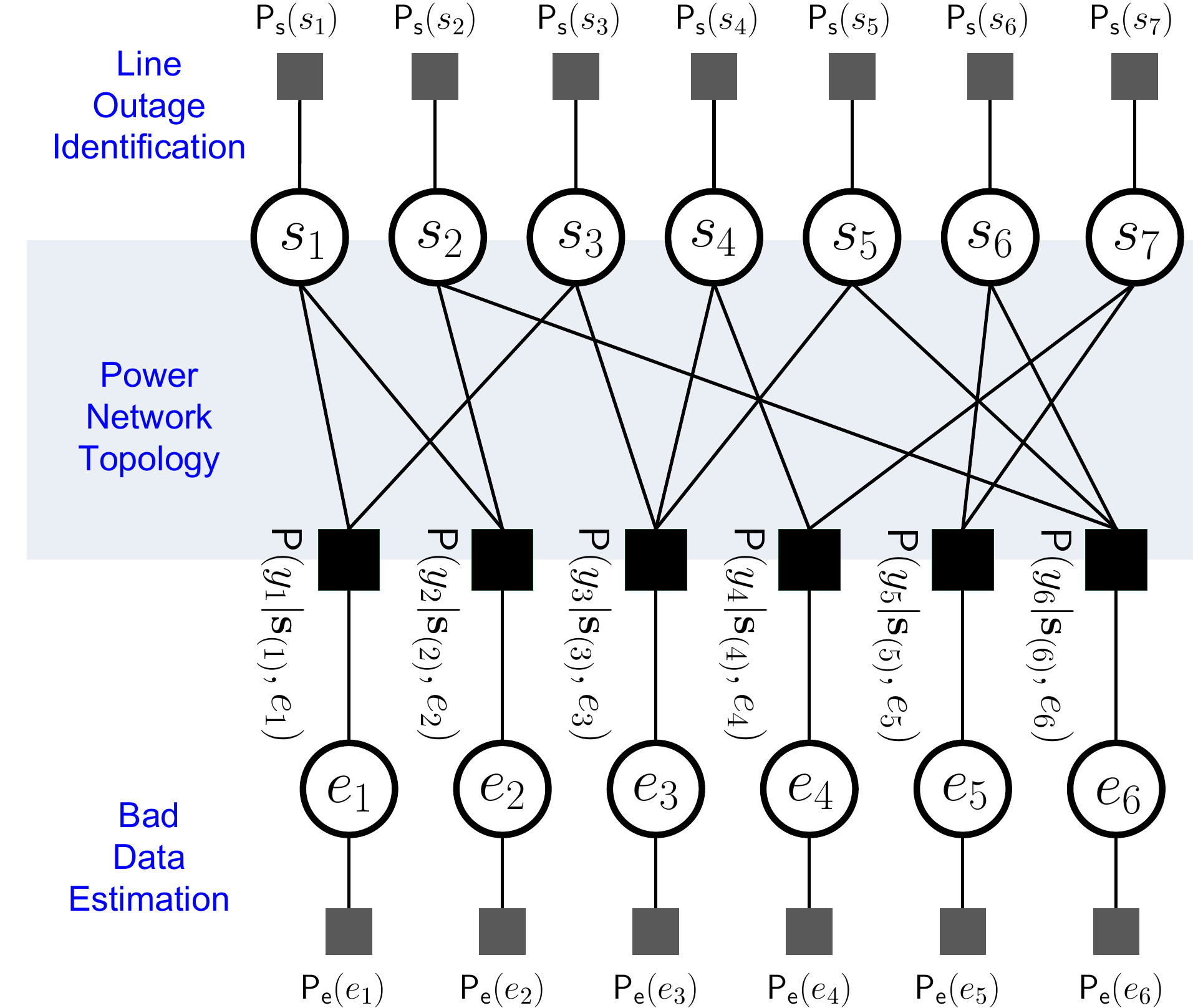} }%
\caption{Factor graph for the six-bus system in Figure \ref{fig:Baddata_example}.}\label{fig:factorgraph_example}
\end{center}
\end{figure}

In this subsection, we develop a computationally efficient algorithm for calculating $\sfQ(s_{l})$ and $\sfQ(e_{n})$, which, in particular, is based on the approximate message passing (AMP) algorithm from \cite{Donoho-09PNAS,Krzakala-12JSM,Manoel-14ArXiv}. For conciseness, we often omit $p_{\sfo}$ from $\sfP_{\sfs}(\qs;p_{\sfo})$ [or $\sfP_{\sfs}(s_{l};p_{\sfo})$], and $\qomega$ from $\sfP_{\sfe}(\qe; \qomega)$ [or $\sfP_{\sfe}(e_{n}; \qomega)$].


AMP can be derived from the perspective of BP, which is a technique to factorize the posterior probability into a product of simpler probability functions. Let $\scrL(n)$ be the set of lines connected bus $n$, and $\scrN(l)$ be the set of buses connected by line $l$. For ease of notation, $[\qs]_{\scrL(n)}$ is denoted by $\qs_{(n)}$. With these notations, the likelihood in (\ref{eq:condProbability}) can be expressed as
\begin{align}
    \sfP(\qy|\qs,\qe)
    &= \prod_{n=1}^{N} \sfP(y_n|\qs_{(n)},e_n) \notag \\
    &=  \frac{1}{\sfZ} \prod_{n=1}^{N}
      e^{-\frac{1}{2 \sigma_{\eta}^2 } \Big\|y_n - \sum_{l \in \scrL(n)} A_{n,l} s_l - e_n \Big\|_2^2 },
      \label{eq:condProbability2}
\end{align}
where $\sfZ$ denotes a normalization factor. Substituting (\ref{eq:P_s}), (\ref{eq:P_e}), and (\ref{eq:condProbability2}) into (\ref{eq:postProbability}), we obtain a factor graph representing the factorization of (\ref{eq:postProbability}) as
\begin{equation} \label{eq:FactorPostPro}
    \sfP(\qs,\qe|\qy)
    = \prod_{n=1}^{N} \sfP(y_n|\qs_{(n)},e_n) {\Bigg(\prod_{l \in \scrL(n)} \sfP_{\sfs}(s_l)\Bigg)}\sfP_{\sfe}(e_n).
\end{equation}
The factor graph for the six-bus system shown in Figure \ref{fig:Baddata_example} is depicted in Figure \ref{fig:factorgraph_example} where a circle represents a variable node associated with the indicator of line outages $s_l$ and contamination $e_n$, whereas a square indicates a factor node associated with the sub-constraint function; i.e., $\sfP(y_n|\qs_{(n)},e_n)$ for bus $n$. For each variable $s_l$, there is an edge between a variable node $l$ and a function node $n$ if and only if line $l$ is connected to bus $n$.

\begin{algorithm}[h]\label{ago:agoSwAMP}  \footnotesize
  \caption{{\tt SwAMP} Algorithm}
  \SetKwInOut{Input}{input}
  \SetKwInOut{Output}{output}
  \SetKwInOut{Initialize}{initialize}
  \SetKwProg{Fn}{}{\string:}{}

  \Input{Input $\qA = [A_{n,l}]\in \bbR^{N \times L}$ and $\qy = [y_{n}] \in \bbR^{N}$. }
  \Output{Return $(\widehat{\qs}, \widehat{\qe})$. }
  \BlankLine
  \Initialize{
  $\widehat{s}_{l}^{0} = 1,~v_{\rms,l}^{0}=N/L,~\forall l$, $\widehat{e}_{n}^{0} = 0,~v_{\rme,n}^{0}=1/N,~\forall n$, $V_{n}^{0} = 1,~\omega_{n}^{0}=y_{n},~\forall n $ }
  \nl $t \leftarrow 1$\;
  \nl \While{ $\sum_{l} (\widehat{s}_{l}^{t} - \widehat{s}_{l}^{t-1})^2 > \epsilon$ {\rm \bf and} $t < T_{\max}$}{
  \nl $g_{n}^{t} \leftarrow \frac{y_{n}-\omega_{n}^{t}}{\sigma_{\eta}^2+V_{n}^{t}}, ~~\forall n $\;
  \nl $V_{n}^{t+1} \leftarrow \sum_{j \in \scrL(n) } A_{n,j}^2 v_{\rms,j}^{t} + v_{\rme,n}^{t}, ~~\forall n$\;
  \nl $\omega_{n}^{t+1} \leftarrow \sum_{j \in \scrL(n)} A_{n,j} \widehat{s}_{j}^{t} + \widehat{e}_{n}^{t} - g_{n}^{t} V_{n}^{t+1}, ~~\forall n$\;

  \BlankLine
  \nl $[\ell_1, \ell_2, \cdots, \ell_{L+N} ] \leftarrow {\tt permute}(\{1,2,\cdots,(L+N)\})$ \;
  \BlankLine
  \nl \For{j = 1 \KwTo L+N}{
  \nl $l \leftarrow \ell_{j}$ \;
  \nl \If{$l \leq L $}{
    \nl $(\Sigma_{\rms,l}^2)^{t+1} \leftarrow  \left( \sum_{i \in \scrN(l)} \frac{|A_{i,l}|^2}{\sigma_{\eta}^2 + V_{i}^{t+1}} \right)^{-1}$\;
    \nl $R_{\rms,l}^{t+1} \leftarrow \widehat{s}_{l}^{t} + (\Sigma_{\rms,l}^2)^{t+1} \sum_{i \in \scrN(l)}
        \frac{A_{i,l} \left(y_{i}-\omega_{i}^{t+1}\right)}{ \sigma_{\eta}^2 + V_{i}^{t+1} }$\;
    \nl $\widehat{s}_{l}^{t+1} \leftarrow f_{\rms,1} \left( (\Sigma_{\rms,l}^2)^{t+1}, R_{\rms,l}^{t+1} \right)$\;
    \nl $v_{\rms,n}^{t+1} \leftarrow f_{\rms,2} \left( (\Sigma_{\rms,l}^2)^{t+1}, R_{\rms,l}^{t+1} \right)$\
        \BlankLine
        \nl\For{$i \in \scrN(l)$ }{
        \nl $\tilde{V}_{i}^{t+1} \leftarrow V_{i}^{t+1}$\;
        \nl $V_{i}^{t+1} \leftarrow V_{i}^{t+1} + A_{i,j}^2 (v_{\rms,j}^{t+1}-v_{\rms,j}^{t}) $\;
        \nl $\omega_{i}^{t+1} \leftarrow \omega_{i}^{t+1}
        +  A_{i,j} (\widehat{s}_{\rms,j}^{t+1} - \widehat{s}_{\rms,j}^{t})
        - g_{i}^{t}( V_{i}^{t+1} - \tilde{V}_{i}^{t+1})$\;
        }
    }
    \BlankLine
  \nl \Else{
    \nl $n \leftarrow l-L$\;
    \nl $(\Sigma_{\rme,n}^2)^{t+1} \leftarrow  \left(\sigma_{\eta}^2 + V_{n}^{t+1}\right)^{-1}$\;
    \nl $R_{\rme,n}^{t+1} \leftarrow \widehat{e}_{n}^{t} + (y_{n}-\omega_{n}^{t+1}) $\;
    \nl $\widehat{e}_{n}^{t+1} \leftarrow {f_{\rme,1} \left( (\Sigma_{\rme,n}^2)^{t+1}, R_{\rme,n}^{t+1} \right)}$\;
    \nl $v_{\rme,n}^{t+1} \leftarrow {f_{\rme,2} \left( (\Sigma_{\rme,n}^2)^{t+1}, R_{\rme,n}^{t+1} \right)}$\;
    \nl $\tilde{V}_{n}^{t+1} \leftarrow V_{n}^{t+1}$\;
    \nl $V_{n}^{t+1} \leftarrow V_{n}^{t+1} + v_{\rme,n}^{t+1}-v_{\rme,n}^{t} $\;
    \nl $\omega_{n}^{t+1} \leftarrow \omega_{n}^{t+1}
    +  (\widehat{e}_{n}^{t+1} - \widehat{e}_{n}^{t})
    - g_{n}^{t}( V_{n}^{t+1} - \tilde{V}_{n}^{t+1})$\;
    }
  }

  \Fn{{\rm \bf Prior parameter learning}}{
  \nl Update the outage probability $p_{\sfo}$ and the B-GM priors $\qomega$ using lines 1--4 in Table \ref{tb:0}\;
  \nl Update the noise variance $\widehat{\sigma}_{\eta}^2$ using line 5 in Table \ref{tb:0}\;
  }

  \nl $t \leftarrow t+1$ \;
  }
\end{algorithm}

In summary, BP can be regarded as a numerically efficient algorithm to obtain (\ref{eq:Qmarginal}) based on the factorization in (\ref{eq:FactorPostPro}). The algorithm is done by a set of message passing equations which go from factor nodes to variable nodes and vice versa. Because of the inherent \emph{sparse} structure of power networks, the computation of the marginal posterior probabilities is tractable using the BP algorithm.

However, the computational complexity is still high because the messages are continuous probabilities. Therefore, we resort to AMP, a variant of BP, which was initially proposed by Donoho {\em et al.} (2009) \cite{Donoho-09PNAS} to solve a linear inverse problem in the context of CS. Applying the AMP technique, we have developed an AMP based algorithm for estimating $(\qs,\qe)$, which is summarized as Algorithm \ref{ago:agoSwAMP}. Wherein, lines 3--5 correspond to the messages from variable nodes $\{ s_{l},e_{n}\}$ to factor nodes, and those lines 9--17 and lines 16--20 correspond to the messages from factor nodes to variable nodes $\{ s_{l} \}$ and variable nodes $\{ e_{n}\}$, respectively. Our version of AMP is closer to \cite{Manoel-14ArXiv}, referred to as the swept AMP ({\tt SwAMP}), which slightly modifies the parallel update patten of AMP to a sequential, or swept, one. We find that {\tt SwAMP} is particularly useful to our case because $\qA$ is a very sparse matrix. In fact, a sparse matrix is a very advantageous situation in terms of computational efficiency. Due to space limitations, we refer the interested reader to \cite{Donoho-09PNAS,Krzakala-12JSM,Manoel-14ArXiv} for more information.

\subsection{Prior Parameter Estimation}
In the above AMP, the prior parameters $(p_{\sfo},\qomega)$ are treated as known. We now apply the EM algorithm in \cite{Vila-13SP} to learn the prior parameters $(\widehat{p}_{\sfo},\widehat{\qomega})$. The EM algorithm is an iterative technique that increases a lower bound on the marginal likelihood $P(\qy; \widehat{p}_{\sfo}^{t}, \widehat{\qomega}^{t})$ at each iteration. Briefly, given a previous parameter estimate $(\widehat{p}_{\sfo}^{t},\widehat{\qomega}^{t})$, the EM update for the parameters is achieved by \cite{Vila-13SP}
\begin{equation} \label{eq:optqq}
\left(\widehat{p}_{\sfo}^{t+1},\widehat{\qomega}^{t+1}\right) = \argmax_{\widehat{p}_{\sfo}^{t},\widehat{\qomega}^{t}} \Ex\left\{ \log \sfP\left(\qy, \qs, \qe; \widehat{p}_{\sfo}^{t},\widehat{\qomega}^{t}\right) \right\},
\end{equation}
where the expectation takes over the posterior probability of $(\qs, \qe)$.

A manipulation for dealing with the optimization (\ref{eq:optqq}) was developed in \cite{Vila-13SP}. Following similar steps in \cite{Vila-13SP}, we can obtain the EM update of the prior parameters $(p_{\sfo},\qomega)$ and the noise level $\sigma_{\eta}^2$, which are summarized as Table \ref{tb:0}. These parameter estimation procedures have been installed in lines 27--28 of Algorithm \ref{ago:agoSwAMP}.

\begin{table}[h]
\begin{center}
\caption{ }\label{tb:0}
\begin{tabular}{|l|}
\hline
 \hspace{1cm} EM update of  prior parameters $(\widehat{p}_{\sfo}^{t+1},\widehat{\qomega}^{t+1})$  \\
\hline \hline
 {\scriptsize 1}~~$\widehat{p}_{\sfo}^{t+1} \leftarrow \frac{1}{L} \sum_{l=1}^{L} \widehat{s}_{l}$,\\
 {\scriptsize 2}~~$\widehat{\rho}_{k}^{t+1} \leftarrow \frac{\sum_{n=1}^{N}\varrho_{k,n}^{t}}{\sum_{n=1}^{N}\overline{\varrho}_{k,n}^{t}}$, \\
 {\scriptsize 3}~~$\widehat{\mu}_k^{t+1} \leftarrow
    \frac{\sum_{n=1}^{N}\varrho_{k,n}^{t} \gamma_{k,n}^{t} }{\sum_{n=1}^{N}\varrho_{k,n}^{t}}$, \\
 {\scriptsize 4}~~$(\widehat{\sigma}_{k}^{2})^{t+1} \leftarrow \frac{\sum_{n=1}^{N}\varrho_{k,n}^{t}\left((\widehat{\mu}_{k,n}-\gamma_{k,n}^{t})^2 + \zeta_{k,n}^{t} \right)}{\sum_{n=1}^{N}\varrho_{k,n}^{t}}$,
 \vspace{-0.25cm} \\ \\ for $k=1,\cdots,K$, where \vspace{-0.25cm} \\  \\
 $\psi_{0,n}^{t} = \rho_{0} \calN\left(\widehat{e}_{n}^{t};  0, v_{\rme,n}^{t} \right)$,~~~~
 $\psi_{k,n}^{t} = \rho_{k} \calN\left(\widehat{e}_{n}^{t}; \widehat{\mu}_{k}^{t}, (\widehat{\sigma}_{k}^{t})^2 + v_{\rme,n}^{t} \right)$,\\
 $\varrho_{k,n}^{t} = \frac{\psi_{k,n}^{t}}{\psi_{0,n}^{t} + \sum_{k'=1}^{K} \psi_{k',n}^{t} }$, ~~
 $\overline{\varrho}_{k,n}^{t} = \frac{\sum_{k'=1}^{K} \psi_{k',n}^{t}}{\psi_{0,n}^{t} + \sum_{k'=1}^{K} \psi_{k',n}^{t} }$, \\
 $\gamma_{k,n}^{t} = \frac{\widehat{e}_{n}^{t}/(\widehat{\sigma}_{k}^{t})^2 + \widehat{\mu}_{k}/v_{\rme,n}^{t}}{1/(\widehat{\sigma}_{k}^{t})^2+1/v_{\rme,n}^{t}}$,  ~~~
 $\zeta_{k,n}^{t} = \frac{1}{1/(\widehat{\sigma}_{k}^{t})^2+1/v_{\rme,n}^{t}}$. \\
\hline \hline
 \hspace{2.0cm} EM update of noise level $\sigma_{\eta}^2$  \\
\hline \hline
 {\scriptsize 5}~~$(\widehat{\sigma}_{\eta}^2)^{t+1} \leftarrow \frac{1}{N}\sum_{n=1}^{N} \left(\frac{|y_{n}-\omega_{n}^{t}|^2/V_{n}^{t}}{1/(\widehat{\sigma}_{\eta}^2)^{t}+1/V_{n}^{t}}+ \frac{1}{1/(\widehat{\sigma}_{\eta}^2)^{t}+1/V_{n}^{t}} \right)$.\\
\hline
\end{tabular}
\end{center}
\end{table}

\section{Simulation Results and Discussion}\label{sec:05}

\begin{table*}
\begin{center}
\caption{(Test Case A) Identification and false alarm rates of various algorithms. }\label{tb:1}
\begin{tabular}{|c|c|c|c|c|c|c|c|c|c|}
  \hline
    \multirow{2}{*}{$N$} &\multirow{2}{*}{$L$}&\multirow{2}{*}{$|\calL_o|$}&Noise&\multicolumn{4}{c|}{Algorithms (Test Case A-1)} & \multicolumn{2}{|c|}{ Alg.~\ref{ago:agoSwAMP} (Test Case A-2) }  \\
    \cline{5-10}
     &&& STD & ~~ES~~  & ~CEO~ & LASSO & Alg.~\ref{ago:agoSwAMP} & ~~~~$\kappa_{\sf I}$~~~~ & $\kappa_{\sf F}$ \\
   \hline\hline
   \multirow{6}{*}{118} & \multirow{6}{*}{179} &\multirow{3}{*}{2}&$0\%$ & $\textbf{100.0}\%$& $99.8\%$& $\emph{99.4}\%$ & $\textbf{100.0}\%$ & $98.4\%$& $0.0\%$ \\
   &&&$1\%$ & $\textbf{99.7}\%$& $99.6\%$& $\emph{98.5}\%$ & $99.6\%$ & $98.4\%$& $0.0\%$ \\
   &&&$3\%$ & $\textbf{98.1}\%$& $97.9\%$& $\emph{95.8}\%$ & $\textbf{98.1}\%$ & $96.2\%$& $0.7\%$\\
   \cline{3-10}
   &&\multirow{3}{*}{3}&$0\%$ & {\rm NA} & $99.8\%$& $\emph{99.3}\%$ & $\textbf{100.0}\%$ & $98.9\%$& $0.0\%$\\
   &&&$1\%$ & {\rm NA} & $99.4\%$& $\emph{98.3}\%$ & $\textbf{99.6}\%$ & $98.7\%$ & $0.3\%$ \\
   &&&$3\%$ & {\rm NA} & $98.0\%$& $\emph{96.2}\%$ & $\textbf{98.2}\%$ & $96.8\%$ & $0.8\%$ \\

  \hline \hline
   \multirow{6}{*}{300} & \multirow{6}{*}{409} &\multirow{3}{*}{2}&$0\%$ & $\textbf{100.0}\%$ & $99.9\%$ & $\emph{99.7}\%$ & $\textbf{100.0}\%$ & $97.6\%$ & $0.0\%$ \\
   &&                  &$1\%$ & $\textbf{98.8}\%$& $98.7\%$ & $\emph{97.5}\%$ & $\textbf{98.8}\%$ & $97.5\%$ & $0.0\%$ \\
   &&                  &$3\%$ & $\textbf{96.9}\%$& $96.8\%$ & $\emph{95.3}\%$ & $\textbf{96.9}\%$ & $95.7\%$ & $1.1\%$ \\
   \cline{3-10}
   &&\multirow{3}{*}{3}&$0\%$ & {\rm NA} & $\textbf{100.0}\%$  & $\emph{99.6}\%$ & $99.9\%$ & $98.2\%$ & $0.0\%$ \\
   &&                  &$1\%$ & {\rm NA} & $99.0\%$ & $\emph{97.8}\%$ & $\textbf{99.1}\%$& $98.1\%$ & $0.0\%$\\
   &&                  &$3\%$ & {\rm NA} & $\textbf{97.3}\%$ & $\emph{95.6}\%$ & $97.2\%$ & $96.0\%$ & $1.0\%$ \\
  \hline \hline
  \multirow{3}{*}{2736} & \multirow{3}{*}{3495} & \multirow{3}{*}{3} & $0\%$ & {\rm NA}& $\textbf{99.9}\%$& $\emph{99.7}\%$ & $\textbf{99.9}\%$  & $97.8\%$ & $0.0\%$ \\
   &&&$1\%$ & {\rm NA}& $\textbf{90.9}\%$& $\emph{90.1}\%$ & $90.5\%$ & $88.2\%$ & $6.3\%$\\
   &&&$3\%$ & {\rm NA}& $\textbf{77.3}\%$& $\emph{76.0}\%$ & $77.1\%$ & $74.9\%$ & $9.1\%$ \\
  \hline
\end{tabular}
\end{center}
\end{table*}

In this section, we conduct computer simulations to demonstrate the
effectiveness and efficiency of the proposed line outage
identification algorithm. Three typical IEEE benchmark power
systems: IEEE $118$-bus, IEEE $300$-bus, and Polish $2736$-bus, are
considered.\footnote{Similar to the other line outage
identification schemes, e.g.,
\cite{Tat-08,Tat-09,Sehwail-12,Zhu-12,Chen-TPS14}, the proposed
scheme is also unable to directly distinguish the outage of a
fraction of multiple lines that connect the ``same'' set of buses.
For this reason, we slightly modify the systems (i.e., the
duplicated lines that connect the same pair of buses are merged into
a single line) to exclude this particular scenario.} The software
toolbox MATPOWER \cite{MATPOWER} is used to generate the phasor
angle measurements corresponding to these power systems, as well as
the pertinent power flows. The performance metrics of our interest
are the identification rate (or the hit rate) and the false alarm
rate; specifically, if $\widehat{\calL}_{\sfo}$ denotes the
\emph{estimate} set of the lines in an outage,\footnote{Recall from
(\ref{eq:tilde_B}) that $\calL_{\sfo}$ denotes the set of the lines
in outage.} the identification rate and the false alarm rate are
defined by
\begin{equation}
    {\kappa_{\sf I} = \frac{\big|\calL_{\sfo} \cap \widehat{\calL}_{\sfo} \big|}{\big|\calL_{\sfo} \big|}}~~~\mbox{and}~~~
    {\kappa_{\sf F} = 1 - \frac{\big|\calL_{\sfo} \cap \widehat{\calL}_{\sfo} \big|}{\big|\widehat{\calL}_{\sfo} \big|}},
\end{equation}
respectively. We will consider the two metrics simultaneously; otherwise, it is known that the identification rate can be trivially high with very high false alarm rate. All the performance results (i.e., the rates) shown in this paper are based on $1,000$ randomly selected locations in outage for each number of line outages. Also, $10$ independent noise-perturbed realizations are generated for each selected location in outage, where the standard deviation (STD) of noise is set equal to $0\%$, $1\%$, or $3\%$ of the average pre-event power injection.

\subsection{Test Case A (without Bad Data)}
In the first experiment, we examine the capability of Algorithm \ref{ago:agoSwAMP} for line outage identification when there are \emph{no} bad data present. In this case, Problem \textsf{P2} reduces to Problem \textsf{P1} since only the estimate of $\qs$ is required. Along this setting, we evaluate the corresponding performers under two different prior knowledge of the system-state.

{\bf Test Case A-1}: In the first case, we assume that the number of line outages $|\calL_{\sfo}|$ and the noise variance $\sigma_{\eta}^2$ are available. We briefly refer the priori information to as the (statistical) system-state information (SSI). Note that our proposed method does not require the availability of the SSI because they can be learned as part of the estimation procedure (i.e., lines 27--28 of Algorithm \ref{ago:agoSwAMP}). However, the SSI is required for the comparison schemes: 1) the LASSO scheme \cite{Zhu-12}, 2) the cross-entropy optimization (CEO) scheme \cite{Chen-TPS14}, and 3) the ES (or exhaustive search). For fair comparisons, we assume perfect prior SSI for all the schemes in this experiment. The results of ES serve as the performance benchmark. However, because the overall complexity of  ES is of the order $\mathcal{O}\left(L^{|\calL_{\sfo}|}\right)$, ES is only available at most for $|\calL_{\sfo}|=2$. Note that since the number of line outages is already known, false alarm is meaningless. Therefore, we only consider the identification rate $\kappa_{\sf I}$ in this case. In addition, note that both the output indicator vectors $\widehat{\qs}_{\sfo}$ by the LASSO and the {\tt SwAMP} are real vectors. These real values can be interpreted as the outage probabilities. Since the number of line outages is known to be $|\calL_{\sfo}|$, we select line outages from the first $|\calL_{\sfo}|$ largest probabilities.

The identification rates of various algorithms with prior SSI are
listed in Table \ref{tb:1} (the first four columns after Noise STD).
As can be seen, under the \emph{same} noise level, the results
obviously show that all the identification rates achieved by
Algorithm~\ref{ago:agoSwAMP} are superior to \emph{all} those by
LASSO and very close to those by the CEO and ES. Particularly
for the Polish 2736-bus system, the identification rates of all the
algorithms become delicate as injection noise level increases. Note
that although the computational complexity of the CEO is smaller
than the ES scheme, it is still much  higher than
Algorithm~\ref{ago:agoSwAMP}. These indicate that
Algorithm~\ref{ago:agoSwAMP} is more suitable than the others in
respect of detection reliability and computational efficiency.

{\bf Test Case A-2}: Next, we consider the cases without the prior SSI (i.e., the number of line outages and the noise variance). Because those comparison schemes mentioned in Case A-1 cannot work effectively as the SSI is unavailable, their results are not included in the following experiments. Note that as the number of line outages is unknown, we estimate $\calL_{\sfo}$ via
\begin{equation} \label{eq:hatL_o}
    \widehat{\calL}_{\sfo} \triangleq \left\{ {l \in \scrL}: \widehat{s}_{\sfo,l} \geq \tau \right\},
\end{equation}
where $0 < \tau < 1$ is the critical number. That is, we perform a \emph{hard decision} from the real vector $\widehat{\qs}_{\sfo}$. It is obvious that a lower value of $\tau$ leads to the higher identification rate $\kappa_{\sf I}$ while also incurring the higher false alarm rate $\kappa_{\sf R}$. Therefore, a proper choice of $\tau$ is important. According to our experiments, we find that $\tau = 0.5$ can generally yield good results. The corresponding results are listed in the last two columns of Table \ref{tb:1}. Comparing the identification rates $\kappa_{\sf I}$ of Algorithm~\ref{ago:agoSwAMP} with prior SSI (column 8) and those without prior SSI (column 9), we see that the identification rates are only \emph{slightly} degraded due to the lack of the prior SSI. In addition, only a very low false alarm rates are arisen. Even for the worst case (except for the inherently delicate 2736-bus system), the false alarm rate is only $1.1\%$. These results illustrate that Algorithm \ref{ago:agoSwAMP} in conjunction with (\ref{eq:hatL_o}) provides a highly effective approach for line outage identification even if the priori SSI is unknown. All the following experiments will be tested without priori SSI.

\subsection{Test Case B (with Bad Data)}

\begin{table}
\begin{center}
\caption{(Test Case B) Performances of Algorithm \ref{ago:agoIdAndRePhase} as the buses with bad data and the associated buses with line outages are involved.}\label{tb:3}
\begin{tabular}{|c|c|c|c|c|l|l|}
  \hline
    \multirow{2}{*}{$N$} &\multirow{2}{*}{$L$}&\multirow{2}{*}{$|\calL_{\sfo}|$}&\multirow{2}{*}{$|\calE_{\sfb}|$}&Noise &\multirow{2}{*}{$\qquad\quad\kappa_{\sf I}$}&\multirow{2}{*}{$\qquad\quad\kappa_{\sf F}$}\\
     &&&& STD &&  \\
   \hline\hline
   \multirow{6}{*}{118} & \multirow{6}{*}{179} & \multirow{6}{*}{3} &\multirow{3}{*}{1}
      & $0\%$ &$ 97.4\%$~ $ 97.8\%$ &$ 40.1\%$~~ $ 2.5\%$ \\
   &&&& $1\%$ &$ 96.5\%$~ $ 96.7\%$ &$ 41.8\%$~~ $ 4.3\%$ \\
   &&&& $3\%$ &$ 94.7\%$~ $ 94.7\%$ &$ 42.6\%$~~ $ 6.3\%$ \\
   \cline{4-7}
   &&&\multirow{3}{*}{2}
      & $0\%$ &$ 92.6\%$~ $ 94.8\%$ &$ 57.6\%$~~ $ 8.4\%$ \\
   &&&& $1\%$ &$ 92.6\%$~ $ 93.7\%$ &$ 58.1\%$~ $ 10.1\%$ \\
   &&&& $3\%$ &$ 91.4\%$~ $ 92.1\%$ &$ 59.1\%$~ $ 13.3\%$ \\

  \hline \hline
   \multirow{6}{*}{300} & \multirow{6}{*}{409} & \multirow{6}{*}{3} &\multirow{3}{*}{1}
      & $0\%$ &$ 97.5\%$~ $ 97.7\%$&$ 41.3\%$~~ $ 2.7\%$ \\
   &&&& $1\%$ &$ 96.5\%$~ $ 96.1\%$&$ 40.6\%$~~ $ 4.8\%$ \\
   &&&& $3\%$ &$ 94.2\%$~ $ 93.5\%$&$ 41.9\%$~~ $ 7.7\%$ \\
   \cline{4-7}
   &&&\multirow{3}{*}{2}
      & $0\%$ &$ 94.4\%$~ $ 95.0\%$&$ 55.8\%$~~ $ 7.9\%$ \\
   &&&& $1\%$ &$ 93.9\%$~ $ 93.1\%$&$ 56.9\%$~ $ 10.4\%$ \\
   &&&& $3\%$ &$ 91.8\%$~ $ 90.5\%$&$ 58.1\%$~ $ 14.8\%$ \\

  \hline \hline
   \multirow{3}{*}{2736} & \multirow{3}{*}{3495} & \multirow{3}{*}{3} &\multirow{3}{*}{2}
      & $0\%$ & $ 97.7\%$~ $ 98.3\%$ & $ 57.0\%$~~ $ 3.2\%$ \\
   &&&& $1\%$ & $ 92.6\%$~ $ 81.7\%$ & $ 61.0\%$~ $ 26.3\%$ \\
   &&&& $3\%$ & $ 86.8\%$~ $ 70.1\%$ & $ 63.7\%$~ $ 40.1\%$ \\

  \hline
\end{tabular}
\end{center}
\end{table}

\begin{table}
\begin{center}
\caption{(Test Case B) Performances of Algorithm \ref{ago:agoIdAndRePhase} as the buses with bad data and the associated buses with line outages are completely separated.}\label{tb:4}
\begin{tabular}{|c|c|c|c|c|l|l|}
  \hline
    \multirow{2}{*}{$N$} &\multirow{2}{*}{$L$}&\multirow{2}{*}{$|\calL_{\sfo}|$}&\multirow{2}{*}{$|\calE_{\sfb}|$}&Noise &\multirow{2}{*}{$\qquad\quad\kappa_{\sf I}$}&\multirow{2}{*}{$\qquad\quad\kappa_{\sf F}$}\\
     &&&& STD &&  \\
   \hline\hline
   \multirow{6}{*}{118} & \multirow{6}{*}{179} & \multirow{6}{*}{3} &\multirow{3}{*}{1}
      & $0\%$ &$ 99.1\%$~ $ 99.0\%$ &$ 42.3\%$~~ $ 5.2\%$ \\
   &&&& $1\%$ &$ 98.0\%$~ $ 97.8\%$ &$ 43.3\%$~~ $ 7.1\%$ \\
   &&&& $3\%$ &$ 96.0\%$~ $ 95.8\%$ &$ 43.7\%$~~ $ 9.3\%$ \\
   \cline{4-7}
   &&&\multirow{3}{*}{2}
      & $0\%$ &$ 97.9\%$~ $ 97.9\%$ &$ 58.6\%$~ $ 14.0\%$ \\
   &&&& $1\%$ &$ 97.3\%$~ $ 97.1\%$ &$ 58.6\%$~ $ 14.9\%$ \\
   &&&& $3\%$ &$ 95.3\%$~ $ 95.1\%$ &$ 59.7\%$~ $ 17.8\%$ \\

  \hline \hline
   \multirow{6}{*}{300} & \multirow{6}{*}{409} & \multirow{6}{*}{3} &\multirow{3}{*}{1}
      & $0\%$ &$ 99.0\%$~ $ 98.9\%$&$ 35.4\%$~~ $ 6.6\%$ \\
   &&&& $1\%$ &$ 98.1\%$~ $ 97.9\%$&$ 36.0\%$~~ $ 7.5\%$ \\
   &&&& $3\%$ &$ 95.7\%$~ $ 95.5\%$&$ 37.2\%$~~ $ 9.6\%$ \\
   \cline{4-7}
   &&&\multirow{3}{*}{2}
      & $0\%$ &$ 99.2\%$~ $ 99.1\%$&$ 50.8\%$~ $ 11.4\%$ \\
   &&&& $1\%$ &$ 97.9\%$~ $ 97.7\%$&$ 51.9\%$~ $ 13.7\%$ \\
   &&&& $3\%$ &$ 95.6\%$~ $ 95.4\%$&$ 52.8\%$~ $ 16.8\%$ \\

  \hline \hline
   \multirow{3}{*}{2736} & \multirow{3}{*}{3495} & \multirow{3}{*}{3} &\multirow{3}{*}{2}
      & $0\%$ & $ 94.7\%$~ $ 94.7\%$ & $ 59.8\%$~~ $ 6.8\%$ \\
   &&&& $1\%$ & $ 88.9\%$~ $ 88.9\%$ & $ 63.8\%$~ $ 31.9\%$ \\
   &&&& $3\%$ & $ 76.3\%$~ $ 76.2\%$ & $ 67.5\%$~ $ 44.7\%$ \\

  \hline
\end{tabular}
\end{center}
\end{table}

In the second experiment, we test the proposed algorithm in the scenarios in which bad data are present at some PMUs.
In the simulations, bad data $[\qtheta'_{\sfb}]_{\calN_{\sfb}}$ are generated by the continuous uniform distribution ${\calU\left(-\bar{\theta},\bar{\theta}\right)}$, where $\bar{\theta}$ is determined by $\frac{1}{N}\sum_{n} |\theta_{n}|$ with $\theta_{n}$ being the pre-event phasor angle. As such, bad data being of similar scale as the common phase angles are concealed in the true data, which makes them difficult to be detected by conventional statistical testes. We apply Algorithm \ref{ago:agoIdAndRePhase} which outputs the indicator vector for the line outage $\widehat{\qs}_{\sfo}$ by solving $(\widehat{\qs},\widehat{\qe})$ from Problem \textsf{P2} followed by the {\sf S-phase} and the {\sf R-phase}. Since $\widehat{\qs}$ from Problem \textsf{P2} is a real vector, we transfer it to a binary vector by using the same hard decision technique as that of (\ref{eq:hatL_o}).

To evaluate the proposed algorithm, we considered two different kinds of bad data locations: the buses with bad data and the associated buses with line outages are i) involved or ii) completely separated. The former case is practical (and more challenging) because it is very likely that the associated buses with line outages result in faulty PMUs. Table \ref{tb:3} and Table \ref{tb:4} list the corresponding results for the two cases with $|\calL_{\sfo}| = 3$ for various numbers of bad data $|\calE_{\sfb}| = \{ 1, \, 2\}$. The results of $\kappa_{\sf I}$ and $\kappa_{\sf F}$ contain two columns. The values in the first column correspond to the results after the {\sf S-phase} (the first phase) of Algorithm \ref{ago:agoIdAndRePhase} and the second columns are the final results of Algorithm \ref{ago:agoIdAndRePhase} (i.e., after the {\sf R-phase}). Recall that the {\sf R-phase} is mainly used to eliminate the state uncertainty between the bad data and line outages. We see that the false alarm rate $\kappa_{\sf F}$ can be greatly reduced by the {\sf R-phase}, and the final identification rate $\kappa_{\sf I}$ remains quite reliable. These results illustrate the effectiveness of Algorithm \ref{ago:agoIdAndRePhase} with bad data.

\subsection{Running Time}

Finally, we discuss the complexity of the proposed line outage
identification algorithm. Given that line 2 of Algorithm
\ref{ago:agoIdAndRePhase} (i.e., Algorithm \ref{ago:agoSwAMP})
\emph{dominates} the computational cost, the complexity of the
proposed line outage identification method can be approximately
analyzed based on the total number of multiplications required by
Algorithm \ref{ago:agoSwAMP}, which requires a total of $3 \sum_{n}
|\scrL(n)| + 8N + 24L + 13KN$ multiplications for each iteration.\footnote{Recall that $K$ defined in (\ref{eq:P_en}) indicates the number of different variance in $\qe$ and we use $K=3$.} To
better grasp the complexity of the entire procedure, we summarize
the average running times of Algorithm \ref{ago:agoIdAndRePhase} in
Table \ref{tb:5} for the test cases A and B. Each running time is obtained by averaging over $10,000$ random samples on a $64$-bit Windows $7$ PC equipped with a $3.3$-GHz Intel Core E3-1230 CPU and
$16$GB of memory. In our simulator, Algorithm
\ref{ago:agoIdAndRePhase} is implemented based on MATLAB $2013$b,
wherein line $2$ (i.e., Algorithm \ref{ago:agoSwAMP}) is written in
the C programming language with $\epsilon = 10^{-6}$ and $T_{\max} = 200$. Table \ref{tb:5} shows that the average
running time increases with the system size and slightly increases
as bad data are present. It can be seen that our algorithm is highly
efficient; the whole identification procedure can be completed
within $1$ second even for the large $2736$-bus system.

\begin{table}
\begin{center}
\caption{Average running time of Algorithm \ref{ago:agoIdAndRePhase} (in seconds)}\label{tb:5}
\begin{tabular}{|c|c|c|c|c|c|c|}
  \hline
    \multirow{2}{*}{$N$} &\multirow{2}{*}{$L$}&\multirow{2}{*}{$|\calL_o|$}&Noise& Test Case A & Test Case B \\
     &&& STD & $|\calE_{\sfb}|=0$ & $|\calE_{\sfb}|=2$ \\
   \hline\hline
   118 & 179 & 3 & $3\%$ & $6.70\times10^{-3}$ & $2.35\times10^{-2}$ \\
  \hline \hline
  300 & 409 & 3 & $3\%$ & $1.61\times10^{-2}$& $4.10\times10^{-2}$ \\
  \hline \hline
  2736 & 3495 & 3 & $3\%$ & $6.24\times10^{-1}$ & $7.82\times10^{-1}$ \\
  \hline
\end{tabular}
\end{center}
\end{table}

\section{Conclusion}\label{sec:06}
In this paper, we developed a framework for identifying multiple power line outages based on the PMUs' measurements in the presence of bad data. Conventionally, the locations of line outages and bad data are indistinguishable. Exploiting the property of power network topology, we presented an algorithm to identify the locations of line outage and recover the faulty measurements simultaneously. The algorithm does not require any prior information of the number of line outages and the noise variance. Simulations using benchmark power systems validated the effectiveness and efficiency of the proposed scheme. In particular, we showed that the whole identification procedure can be completed within seconds even for a large-scale power system, which makes our scheme suitable for real-time applications.



\begin{thebibliography}{10}
\providecommand{\url}[1]{#1}
\csname url@samestyle\endcsname
\providecommand{\newblock}{\relax}
\providecommand{\bibinfo}[2]{#2}
\providecommand{\BIBentrySTDinterwordspacing}{\spaceskip=0pt\relax}
\providecommand{\BIBentryALTinterwordstretchfactor}{4}
\providecommand{\BIBentryALTinterwordspacing}{\spaceskip=\fontdimen2\font plus
\BIBentryALTinterwordstretchfactor\fontdimen3\font minus
  \fontdimen4\font\relax}
\providecommand{\BIBforeignlanguage}[2]{{%
\expandafter\ifx\csname l@#1\endcsname\relax
\typeout{** WARNING: IEEEtran.bst: No hyphenation pattern has been}%
\typeout{** loaded for the language `#1'. Using the pattern for}%
\typeout{** the default language instead.}%
\else
\language=\csname l@#1\endcsname
\fi
#2}}
\providecommand{\BIBdecl}{\relax}
\BIBdecl

\bibitem{Aminifar-14Access}
{F. Aminifar, M. Fotuhi-Firuzabad, A. Safdarian, A. Davoudi, and M.
  Shahidehpour}, ``{Synchrophasor measurement technology in power systems:
  panorama and state-of-the-art},'' \emph{IEEE Access}, vol.~2, pp. 1607--1628,
  2014.

\bibitem{Tat-08}
{J.~E.~Tate and T.~J.~Overbye}, ``Line outage detection using phasor angle
  measurements,'' \emph{IEEE Trans. Power Syst.}, vol.~23, no.~4, pp.
  1644--1652, Nov. 2008.

\bibitem{Tat-09}
------, ``Double line outage detection using phasor angle measurements,'' in
  \emph{Proc. IEEE PES Gen. Meeting}, 2009, pp. 1--5.

\bibitem{Sehwail-12}
{H. Sehwail and I. Dobson}, ``{Locating line outages in a specific area of a
  power system with synchrophasor},'' in \emph{2012 North American Power
  Symposium (NAPS)}, Champaign, IL, 9-11 Sep. 2012, pp. 1--6.

\bibitem{Emami-TPS13}
{T. Emami and A. Abur}, ``{External system line outage identification using
  phasor measurement units},'' \emph{IEEE Trans. Power Syst.}, vol.~28, no.~2,
  pp. 1035--1040, May 2013.

\bibitem{He-TSG11}
{M. He and J. Zhang}, ``{A dependency graph approach for fault detection and
  localization towards secure smart grid},'' \emph{IEEE Trans. Smart Grid},
  vol.~2, no.~2, pp. 342--351, Jun. 2011.

\bibitem{Abdelaziz-12}
{A. Y. Abdelaziz, S. F. Mekhamer, M. Ezzat, and E. F. El-Saadany}, ``{Line
  outage detection using support vector machine (SVM) based on the phasor
  measurement units (PMUs) technology},'' in \emph{2012 IEEE Power and Energy
  Society General Meet.}, San Diego, CA, 22-26 July 2012, pp. 1--8.

\bibitem{Zhu-12}
{H.~Zhu and G.~B.~Giannakis}, ``Sparse overcomplete representations for
  efficient identification of power line outages,'' \emph{IEEE Trans. Power
  Syst.}, vol.~27, no.~4, pp. 2215--2224, Nov. 2012.

\bibitem{Chen-TPS14}
{J.-C. Chen, W.-T. Li, C.-K. Wen, J.-H. Teng, and P. Ting}, ``{Efficient
  identification method for power line outages in the smart power grid},''
  \emph{IEEE Trans. Power Syst.}, vol.~29, no.~7, pp. 1788--1800, July 2014.

\bibitem{Zhao-S14}
{L. Zhao and W.-Z. Song}, ``{Distributed power-line outage detection based on
  wide area measurement system},'' \emph{Sensors}, vol.~14, no.~7, pp.
  13\,114--13\,133, Jan. 2014.

\bibitem{Wu-15TPS}
{J. Wu, J. Xiong, and Y. Shi}, ``{Efficient location identification of multiple
  line outages with limited PMUs in smart grids},'' \emph{IEEE Trans. Power
  Syst.}, 2015.

\bibitem{Chen-14JSTSP}
{C. Chen, J. Wang, and H. Zhu}, ``{Effects of phasor measurement uncertainty on
  power line outage detection},'' \emph{IEEE J. Sel. Topics Signal Process.},
  vol.~8, no.~6, pp. 1127--1139, Dec. 2014.

\bibitem{Hayashi-13IEICE}
{K. Hayashi, M. Nagahara, and T. Tanaka}, ``{A user's guide to compressed
  sensing for communications systems},'' \emph{IEICE Trans. Commun.}, vol.
  E96-B, no.~3, pp. 685--712, Mar. 2013.

\bibitem{Wood-96BOOK}
A.~J. Wood and B.~F. Wollenberg, \emph{{Power Generation, Operation and
  Control}}.\hskip 1em plus 0.5em minus 0.4em\relax New York: Wiley, 1996.

\bibitem{Sch-05}
{A.~Schellenberg, W.~Rosehart, and J.~Aguado}, ``Cumulant-based probabilistic
  optimal power flow (p-opf) with gaussian and gamma distributions,''
  \emph{IEEE Trans. Power Syst.}, vol.~20, no.~2, pp. 773--781, May 2005.

\bibitem{Tibshirani-96JRSS}
{R. Tibshirani}, ``{Regression shrinkage and selection via the lasso},''
  \emph{J. Royal. Statist. Soc., Ser. B}, vol.~58, no.~1, pp. 267--288, 1996.

\bibitem{Niazadeh-10}
{R.~Niazadeh, M.~Babaie-Zadeh, and C.~Jutten}, ``An alternating minimization
  method for sparse channel estimation,'' in \emph{in Ninth International
  Conference on Latent Variable Analysis and Signal Seperation}, 2010, pp.
  319--327.

\bibitem{Poor-94BOOK}
H.~V. Poor, \emph{{An Introduction to Signal Detection and Estimation}}.\hskip
  1em plus 0.5em minus 0.4em\relax New York: Springer-Verlag, 1994.

\bibitem{Donoho-09PNAS}
{D. L. Donoho, A. Maleki, and A. Montanari}, ``{Message passing algorithms for
  compressed sensing},'' \emph{Proc. Nat. Acad. Sci.}, vol. 106, no.~45, pp.
  18\,914--18\,919, 2009.

\bibitem{Krzakala-12JSM}
{F. Krzakala, M. M\'{e}zard, F. Sausset, Y. Sun, and L. Zdeborov\'{a}},
  ``{Probabilistic reconstruction in compressed sensing: algorithms, phase
  diagrams, and threshold achieving matrices},'' \emph{J. Stat. Mech.}, vol.
  P08009, 2012.

\bibitem{Vila-13SP}
{J. P. Vila and P. Schniter}, ``{Expectation-maximization Gaussian-mixture
  approximate message passing},'' \emph{IEEE Trans. Sig. Proc.}, vol.~61,
  no.~19, pp. 4658--4672, Oct. 2013.

\bibitem{Manoel-14ArXiv}
{A. Manoel, F. Krzakala, E. W. Tramel, L. Zdeborov\'{a}}, ``{Sparse estimation
  with the swept approximated message-passing algorithm},'' \emph{arXiv
  preprint 1406.4311}.

\bibitem{MATPOWER}
{R.~D.~Zimmerman, C.~E.~Murillo-S\'{a}nchez, and R.~J.~Thomas}, ``{MATPOWER}
  steady-state operations, planning and analysis tools for power systems
  research and education,'' \emph{IEEE Trans. Power Syst.}, vol.~26, no.~1, pp.
  12--19, Feb. 2011.

\end{thebibliography}
\end{document}